\documentclass [aps,prl,twocolumn,groupeauthor,showpacs,superscriptaddress]{revtex4}
\usepackage{amsmath,amssymb,latexsym}

\usepackage{graphicx}
\usepackage{bm}

\begin{document}

\title{Ultrafast separation of photo-doped carriers in Mott antiferromagnets}

\author{Martin~Eckstein}
\affiliation{Max Planck Research Department for Structural Dynamics, University of Hamburg-CFEL, 22761 Hamburg, Germany}
\email{martin.eckstein@mpsd.cfel.de}
\author{Philipp~Werner}
\affiliation{Department of Physics, University of Fribourg, 1700 Fribourg, Switzerland}
\begin{abstract}
We use inhomogeneous nonequilibrium dynamical mean-field theory to investigate the spreading of photo-excited carriers in Mott insulating heterostructures with strong internal fields. Antiferromagnetic correlations are found to affect the carrier dynamics in a crucial manner: An antiferromagnetic spin background can absorb energy from photo-excited carriers on an ultrafast 
timescale, thus enabling fast transport between different layers and the separation of electron and hole-like 
carriers, whereas in the paramagnetic state, carriers become localized in strong fields. 
This interplay between charge and spin degrees of freedom can be exploited to control the functionality 
of devices based on Mott insulating heterostructures with polar layers, e.g., for photovoltaic applications.
\end{abstract}
\date{\today}

\maketitle

Heterostructures and interfaces of transition-metal oxides \cite{Ohtomo2002,Ohtomo2004,Okamoto2004}  allow to exploit correlation effects  on atomic length-scales for the 
purpose of material design, while femtosecond laser pulses can probe and manipulate the evolution of complex states of matter such as superconducting \cite{Fausti11} or 
magnetically ordered phases \cite{Kirilyuk10} on microscopic timescales. Combining ideas and techniques from both fields \cite{Caviglia2012} promises an unprecedented 
level of control over complex dynamical processes. In this paper we analyze in particular the intriguing nonlinear carrier dynamics in antiferromagnetic Mott-insulating 
heterostructures which results from the interplay of spin and charge degrees of freedom on femtosecond timescales. Recent theoretical work on non-linear transport in 
correlated systems \cite{Mierzejewski2011,Vidmar2011,Amaricci2012,Golez2013} 
has revealed a somewhat counter-intuitive effect of dissipation on the carrier dynamics in strong fields. In a closed
system, particles remain localized in a potential energy gradient, because the electronic bandwidth limits their kinetic
energy. Transport is thus enhanced, rather than hampered, by dissipation and the dominant scattering channel, 
such as phonons or magnons, defines the maximum possible current \cite{Mierzejewski2011,Vidmar2011}. Related 
nonlinearities can be expected to affect the transport properties of heterostructures with polar layers, which can 
build up internal fields of order $0.1$ Volt per lattice spacing \cite{Nakagawa2006,SinghBhalla2011}. 

An interesting device where the carrier separation by strong internal fields plays an important role is the Mott insulating solar cell 
proposed in  Ref.~\onlinecite{Assmann2013}.  Its efficient operation is possible only if one can beat the rapid 
recombination of photo-excited carriers, which happens on much shorter (picosecond) timescales in Mott systems 
than in semiconductors \cite{Okamoto2010,Mitrano2013}. As we will demonstrate, an ultra-fast separation of 
photo-induced carriers is indeed possible in antiferromagnetically ordered materials, where it relies on the 
capability of the spin sector to act as an ``energy bank" which can absorb and store energy on femtosecond 
timescales, before slowly dissipating it to the lattice and/or leads in a manner which is independent of the 
dynamics of the photo-doped carriers. In a spin disordered state, on the other hand, the estimated carrier 
separation times become comparable to typical carrier recombination times in Mott insulators,  even for fields 
approaching the electric breakdown limit of the device. Our results also suggest a way to use spatial inhomogeneities 
to explore the ultrafast scattering between spins and carriers, complementary to optical spectroscopy \cite{DalConte2012}.

\begin{figure}[tbp]
\includegraphics[width=0.92\columnwidth,]{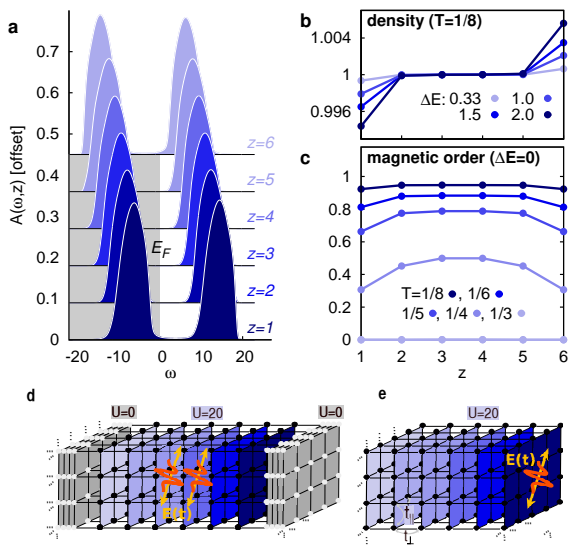}
\caption{
Equilibrium configuration of the heterostructure and nonequilibrium setup.
(a)  
Layer and frequency resolved spectral function $A(\omega,z)$ in the 
Mott heterostructure with leads, for temperature $T=1/3$ (paramagnetic phase)
and internal field $\Delta E=1.5$; $\omega=0$ is the Fermi level of 
the external leads.
(b)
Layer-dependent density for various gradients $\Delta E$.
(c)
Layer-dependent antiferromagnetic order for various temperatures up 
to the N\'eel temperature. 
(d) Setup used to study the doublon and hole diffusion to the leads.
(e) Setup without leads, which allows to study the spreading of doublons over a longer distance. 
\label{fig1}}
\end{figure}

We demonstrate the nontrivial effect of the spin background on the carrier dynamics by considering a model 
consisting of six layers of a $3$-dimensional Mott insulator with simple cubic lattice structure, coupled to 
metallic leads (Fig.~\ref{fig1}(d)). The insulator is described by a Hubbard Hamiltonian,
\begin{align}
H
=&
\sum_{ z, \langle \bm i\bm j \rangle, \sigma} t_{ \bm i  \bm j}^{|\!|}
c_{\bm i,z,\sigma}^\dagger
c_{\bm j,z,\sigma}
+
\sum_{\langle z,z'\rangle,  \bm i,  \sigma} t^\perp_{z,z'}
c_{\bm i,z,\sigma}^\dagger
c_{\bm i,z',\sigma}
\nonumber
\\
&+U
\sum_{\bm j,z}
n_{\bm j,z,\uparrow}n_{\bm j,z,\downarrow}
-\Delta E
\sum_{\bm j,z,\sigma}
(z-3.5)\, n_{\bm j ,z,\sigma}\,,
\label{repulsive hubbard}
\end{align}
with local Coulomb repulsion $U$, nearest-neighbor intra-layer hopping $t_{ \bm i  \bm j}^{|\!|}$, 
and inter-layer hopping $t^\perp_{z,z'}$; $z=1,...,6$ denotes the layer, and $\bm i$ is the lattice 
coordinate within the infinitely extended $x$-$y$ plane.  A static potential energy difference $\Delta E$ 
between the layers captures the effect of internal fields in the polar structure. 

{\bf Methods --} To solve the electron dynamics, we use nonequilibrium dynamical mean-field theory (DMFT) 
\cite{Freericks2006,REVIEW}. DMFT 
maps a correlated lattice problem onto a self-consistently determined impurity model \cite{Georges1996}. 
The main approximation in this formalism is the assumption of a spatially local self-energy. While spatial 
correlations are treated at a mean-field level, time-dependent fluctuations can be captured accurately. 
Hence, the formalism is suitable for the study of nonequilibrium phenomena in lattice models. Inhomogeneous 
layered structures can be treated via a mapping to a coupled set of impurity problems, each representing 
one layer \cite{Potthoff1999,Freericks2004}. A detailed description of the formalism and of our numerical 
implementation is given in  Ref.~\onlinecite{REVIEW} 
for the translationally invariant case, and in  Ref.~\onlinecite{Eckstein2013layer} for the inhomogeneous DMFT 
formalism. To take into account antiferromagnetic order, we have implemented this formalism with a magnetic 
unit cell of two sites within the layers. Metallic leads with a flat density of states are included in the 
formalism by adding additional lead self-energies \cite{Eckstein2013layer}.
The impurity model is solved here using the perturbative hybridization 
expansion (non-crossing approximation, NCA) \cite{Eckstein2010nca}. The validity of the NCA for the (photo-doped) 
Mott insulating phase has been addressed previously for homogeneous bulk systems. NCA gives qualitatively 
the same results as higher order variants of the expansion for the current in the dielectric breakdown regime 
\cite{Eckstein10} and for the relaxation of a photo-excited Mott-insulator \cite{Eckstein11}. Simulations of  
inhomogeneous systems in an antiferromagnetic state are numerically considerably more demanding, 
such that higher-order schemes (one-crossing approximation) are not a realistic option at this time. 

We compute the time-evolution of various observables, in particular the (layer-resolved) density 
$n(t,z) = (1/L) \sum_{\bm j\sigma} \langle n_{\bm j,z,\sigma}\rangle $ and staggered magnetization 
$m(t,z) = (1/L) \sum_{\bm j} e^{i \bm q \bm j}\langle n_{\bm j,z,\uparrow}-n_{\bm j,z,\downarrow}\rangle $, with $\bm q=(\pi,\pi)$.
The photoemission spectrum  $I(\Omega,t_p,z)$ is obtained from a convolution of the lesser 
component of the local, layer-resolved Green's function $G_{z}^<(t,t') = i\sum_\sigma
\langle c_{\bm i z\sigma}^\dagger(t') c_{\bm i z\sigma}(t)\rangle $
with the probe-pulse envelope $s(t)=\exp(-t^2/\delta t^2)$ 
\cite{FreericksKrishnamurthyPruschke2009},
\begin{equation}
I(\Omega,t_p,z) 
\!=\!\!\!
\int \!dt dt' s(t-t_p)s(t'-t_p) e^{i\Omega(t-t')} \,G_{z}^<(t,t').
\end{equation}
The width $\delta t$ and 
its inverse $\hbar/\delta t$ determine the time and energy resolution, respectively.
The spectrum also provides an unambiguous way to measure the density of photo-excited 
doublon and hole carriers, $n_{d}(t,z)$ and $n_h(t,z)$: 
Physically, a doublon is a 
resonance that is well defined only on timescales much longer than that of virtual charge 
fluctuations, an hence $n_{d}(t,z)$ and $n_h(t,z)$ are given by the integrated photoemission 
weight in the upper Hubbard band and the integrated inverse photoemission weight in the 
lower Hubbard band, respectively. (In a nonequilibrium state with simultaneous 
excitation of doublons and holes, carrier densities cannot be computed from the density.)

\begin{figure}[tbp]
\includegraphics[width=0.92\columnwidth]{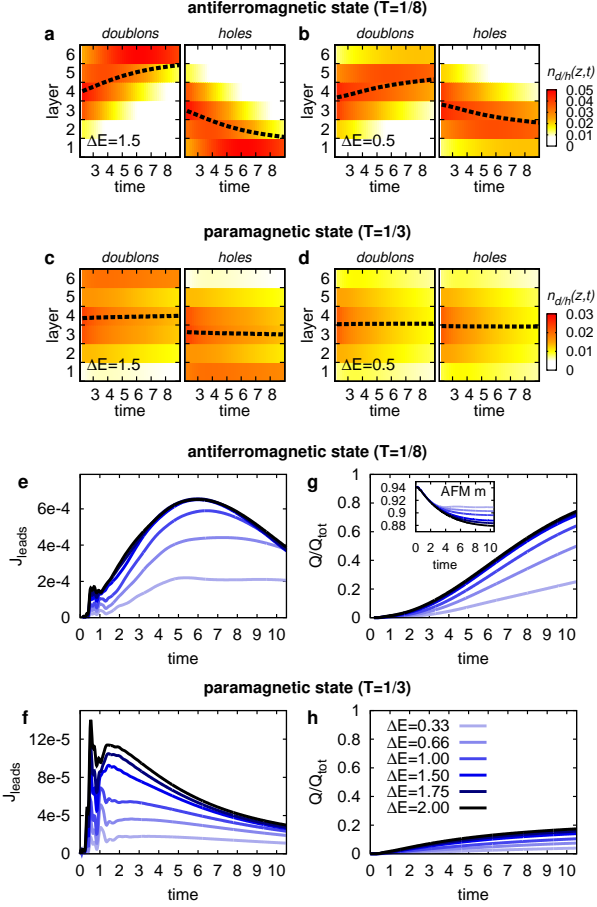}
\caption{
(a)-(d) Time and layer-dependent doublon and hole-densities for two different internal fields.  
The dashed lines indicate the mean $\overline z_{d/h}(t)$ of the distributions in space. 
The heterostructure 
is excited at layers $z=3,4$.
(e),(f) Current to the leads for various $\Delta E$ 
(see key in panel (h)). Here, the heterostructure is excited at all layers. (g),(h) 
Time-integrated current $Q=\int_{0}^t dt' J_\text{lead}(t')$, compared to the total photo-excited 
charge, for the same parameters as in (e) and (f). The inset in (g) shows the layer-averaged
antiferromagnetic order parameter as a function of time.
\label{fig2}}
\end{figure}

{\bf Results --}  We choose $|t^{|\!|}|=|t^{\perp}|\equiv t_*$ as the unit of energy, and take $U=20$, such that 
the system becomes a  Mott insulator with a gap comparable to the free $3$-dimensional bandwidth 
$12 t_*$. For a gap of $1$eV \cite{Assmann2013}, the unit of time is thus $\hbar/t_*\approx 5$fs. 
In Figure~\ref{fig1}(a) we plot the layer-resolved spectral functions 
of the heterostructure in equilibrium. All simulations are performed for un-biased leads. Up to $\Delta E\approx 2$, 
where the Hubbard bands at the boundary become degenerate with the Fermi energy $E_F=0$, the coupling to the 
noninteracting leads yields only a weak transfer of spectral weight into the gap region at the boundary layers, and 
slight hole/electron doping (Fig.~\ref{fig1}(b)). The system becomes antiferromagnetically ordered for temperatures 
below $T=T_N\approx0.3$ (Fig.~\ref{fig1}(c)). 

The photo-excitation is simulated by a few-cycle in-plane electric field pulse with frequency 
$\Omega=U$, $\bm E(t)=(E_0,E_0,0) \sin(\Omega t) \exp(-4.605 (t-t_0)^2/ t_0^2)\Theta(2t_0-t) \Theta(t)$, 
centered around time $t_0=2\pi/\Omega$ (the field is included as  a Peierls phase of the intra-layer
hopping \cite{Eckstein2013layer}).
The pulse promptly creates a small density of hole- and electron-like carriers 
(doublons) in the Mott insulating structure. As expected for $U/t_* \gg 1$, we 
observe no recombination of photo-excited carriers within the timescale of our simulation 
\cite{Sensarma2010a,Eckstein11}. The evolution of the doublon and hole densities $n_{d}(t,z)$ 
and $n_{h}(t,z)$ under the influence of the internal field is fundamentally different in the 
paramagnetic and antiferromagnetic structures (Fig.~\ref{fig2}(a)-(d)): In the latter, doublons 
and holes rapidly separate and move to the boundary layers, where they 
are collected by the leads. This process becomes faster for larger $\Delta E$, as evidenced 
by the spatial mean $\overline z_{d/h}(t) = \sum_{z}z\,n_{d/h}(z,t)/\sum_{z} n_{d/h}(z,t)$ 
of the distributions (dotted lines in Figs.~\ref{fig2}(a)-(d)). In the paramagnetic phase, on the 
other hand, we observe almost no separation of carriers beyond a polarization proportional 
to $\Delta E$, which builds up already during the excitation (offset between $\overline  z_{d}$ 
and $\overline  z_{h}$). The current $J_\text{lead}$ into the leads shows a similar tendency 
(Fig.~\ref{fig2}(e)-(h)). In the antiferromagnetic sample, it reaches its maximum roughly when 
$50\%$ of the total number of photo-excited carriers have tunneled into the leads, which 
takes only a few hopping times for large fields $\Delta E \gtrsim 1$. In the paramagnet, 
the current starts to decrease already shortly after the pulse, when most of the photo-carriers 
still reside within the heterostructure. The simulations are limited to short times, but 
we can take the current at $t=10$ to estimate a lower bound for the time needed for the 
carriers to leave the structure. At $\Delta E=1$, this gives $t=150$, i.e., $0.75$ps for a gap of $1$eV. 
On picosecond timescales, however, weak interactions such as couplings to optical phonons 
or  nonlocal Coulomb interactions not included in our model Hamiltonian may lead to exciton 
formation and recombination \cite{Lenarcic2013,Mitrano2013}, which would degrade the 
efficiency of the device.

\begin{figure}[tbp]
\includegraphics[width=0.92\columnwidth]{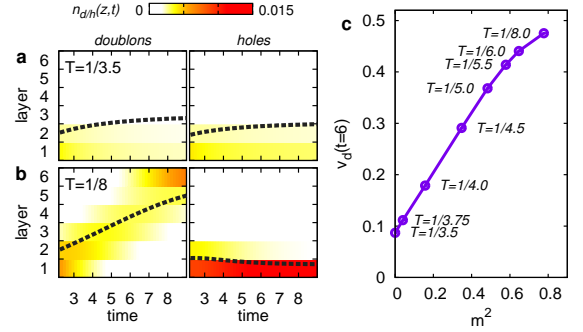}
\caption{
Temperature dependence of carrier motion.
(a),(b) Layer and time-resolved doublon and hole densities for two temperatures, $T=1/3.5$ 
(paramagnetic) and $T=1/8$ (antiferromagnetic). To study the carrier motion over a longer distance, 
we consider a heterostructure which is not coupled to leads, apply the excitation at layer $z=1$, 
and focus on the doublon dynamics  (holes remain trapped at the boundary), see sketch in Fig.~\ref{fig1}(e).  The internal field is 
$\Delta E=1$. The dotted line shows the spatial mean $\overline z_{d/h}(t)$. (c) Velocity 
$v_d(t)=d \overline z_{d}(t)/dt$ at intermediate time $t=6$, plotted for various temperatures against the square 
$m^2$ of the layer-averaged staggered magnetization.
\label{fig3}}
\end{figure}

To explain the striking difference between paramagnetic and antiferromagnetic samples, we 
first note that the decaying current in the paramagnetic sample resembles the behavior of 
isolated (nonintegrable) metals \cite{Mierzejewski2010,Eckstein2011bloch,Amaricci2012} 
and Mott insulators \cite{Eckstein10}. In fact, to separate doublons and holes by one layer, 
the charges must release an energy proportional to $\Delta E$. Heat transport to the leads 
takes place, but turns out to be slow. In the antiferromagnetic phase, on the other hand, 
flipping spins provides a fast channel to release this energy {\em locally}. Indirect evidence for 
an energy transfer to the spin sector is given by the decrease of the magnetization $m$ 
with time (inset in Fig.~\ref{fig2}(g)). Furthermore, we find that the drift velocity $v_d(t) =  
d/dt\,\overline z_d(t)$ increases with $m^2$ as the temperature is decreased and the 
magnetic order strengthens (Fig.~\ref{fig3}). Such a behavior may be expected because 
for each hopping the average number of spin-flips against the antiferromagnetic exchange 
$J=4t_*^2/U$ is proportional to $m$, and hence the maximal rate of change of the spin 
energy $Jm^2$ is proportional to $m^2$ and a hopping rate.

\begin{figure}[tbp]
\includegraphics[width=0.92\columnwidth]{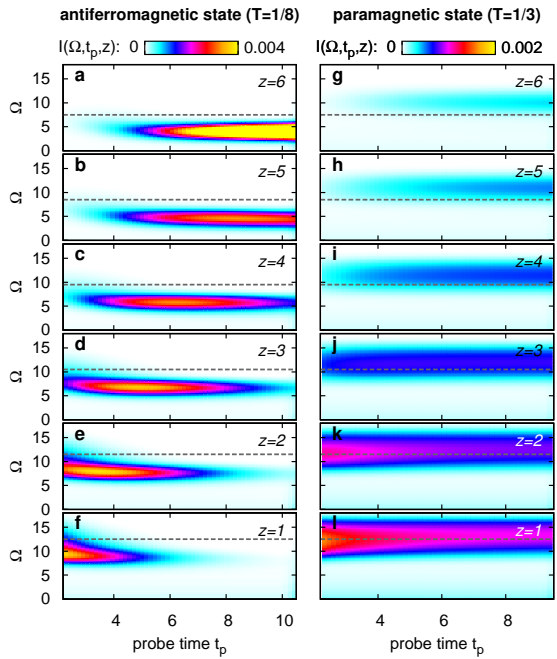}
\caption{
Layer-resolved, spin-integrated photoemission spectrum $I(\Omega,t_p,z)$ of the upper 
Hubbard band, for the same geometry as in Fig.~\ref{fig3}.
(a)-(f) Antiferromagnetic structure at $T=1/8$.  (g)-(l) Paramagnetic structure 
at  $T=1/3$. The dashed lines indicate the center of the upper Hubbard band.
\label{fig4}}
\end{figure}

A more detailed picture of the  carrier dynamics can be obtained from the photoemission 
spectrum $I(\Omega,t_p,z)$ (occupied density of states) as a function of layer, probe time 
$t_p$ and frequency (Fig.~\ref{fig4}). We computed $I(\Omega,t_p,z)$ with a time resolution 
of $\delta t=1$  (set by the probe duration), and an uncertainty-limited frequency resolution 
$\delta \omega=\hbar/\delta t$ (see methods). In the antiferromagnetic case, the initially broad 
energy distribution in the upper Hubbard band of the excited layer ($z=1$) rapidly accumulates 
at the lower band edge, which demonstrates the transfer of kinetic energy of photo-carriers to 
the spin background. This cooling mechanism is absent in the paramagnetic phase, and 
consequently, there is no significant redistribution of spectral weight within the band,
as it has also been observed in homogeneous Mott systems \cite{Eckstein11,Moritz2012}.
As the carriers gradually move towards the other boundary, the spectral weight in the 
antiferromagnetic structure further accumulates in the lower half of the Hubbard band, 
while it remains pinned at the same energy in the paramagnetic calculation (thus shifting 
upwards within the Hubbard band layer by layer).

\begin{figure}[tbp]
\includegraphics[width=0.94\columnwidth]{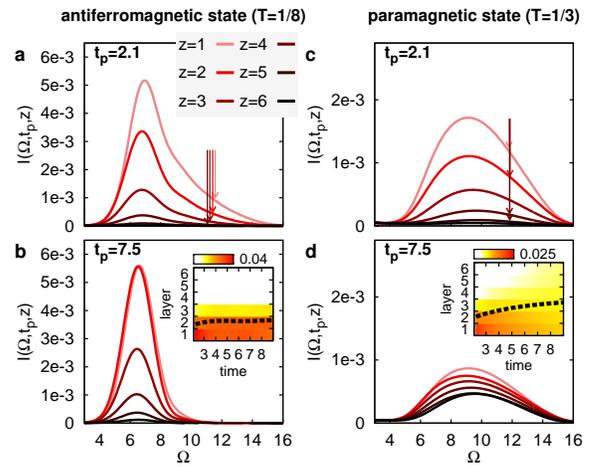}
\caption{
Diffusion of carriers in zero field: Photoemission spectrum of the 
upper Hubbard band shortly after the excitation ($t_p=2.1$) and at time
$t_p=7.5$. The setup is identical to Fig.~\ref{fig3}, but 
$\Delta E=0$. 
(a) $t_p=2.1, T=1/8$.
(b) $t_p=7.5, T=1/8$.
(c) $t_p=2.1, T=1/3$.
(d) $t_p=7.5, T=1/3$.
Vertical arrows in (a) and (c) indicate the mean of the distribution 
in the spectral region $10<\omega<16$. Insets in (b) and (d): 
density $n_{d}(z,t)$ for the respective temperatures.
 \label{fig5}}
\end{figure}

Interestingly, the opposite effect is observed for the pure diffusion of carriers at 
$\Delta E=0$ \cite{Golez2013}. Diffusing carriers in the antiferromagnetic phase need kinetic energy 
to perturb the magnetic order. After a short time, the spectral occupation is concentrated 
at the lower edge of the upper Hubbard band, and further diffusion relies on much 
slower energy transport processes (Fig.~\ref{fig5}(a),(b)). The high-energy tails of the 
distributions at early times are layer-dependent, which reveals the energy loss of 
carriers as they move from layer to layer. In the paramagnet, on the other hand, 
doublons and holes spread throughout the sample in a purely diffusive manner, 
and the distribution functions remain broad at all layers and times (Fig.~\ref{fig5}(c),(d)).

In summary, our results demonstrate the decisive role of spin flip scattering for the 
carrier dynamics in Mott heterostructures with antiferromagnetic correlations. The 
spin background can act as a buffer which absorbs energy on fs timescales, and thus
enables a rapid carrier separation in photovoltaic devices (the LaVO$_3$ system 
proposed in  Ref.~\onlinecite{Assmann2013} is indeed magnetically ordered).
In a broader context, exploiting the spin scattering and its ability to rapidly cool injected 
carriers may become an important design principle for devices which operate on a fs timescale.
It can boost the transport of carriers in the strong field gradients of polar heterostructures, 
or significantly delay the spreading of carriers in zero field. 
Conversely, inhomogeneous set-ups may be used to probe the ultra-fast exchange of 
energy between particles and spins: While optical excitations typically penetrate 
many atomic layers, photoemission can be made surface sensitive. In a 
heterostructure with band bending towards the surface, one can thus expect 
a delay between the maximum photoemission signal and the pump,  which 
reflects the spreading of the carriers.

\section*{Acknowledgments}
We thank J. Bon\v{c}a, K. Held, D. Gole\v{z}, Z.~Lenar\v{c}i\v{c}, T.~Oka, P. Prelov\'sek, G.~Sangiovanni, N. Tsuji, and L.~Vidmar  for 
fruitful discussions. The calculations were run on the supercomputer HLRN-II of the North-German 
Supercomputing Alliance. PW acknowledges support from FP7/ERC starting grant No. 278023.


\end{document}